\documentclass[conference,a4paper]{format}
\usepackage{amsmath}
\usepackage{graphicx}
\usepackage{multirow}
\usepackage{threeparttable}
\usepackage[backend=biber,style=ieee,]{biblatex}
\usepackage{threeparttable}
\usepackage{booktabs}
\usepackage{amssymb}
\usepackage[hidelinks]{hyperref}
\usepackage{url}

\usepackage{geometry}
\usepackage{fancyhdr}

\begin{document}

\title{A Human-Inspired Decoupled Architecture for Efficient Audio Representation Learning}

\author{
\authorblockN{
Harunori Kawano\authorrefmark{1} and
Takeshi Sasaki\authorrefmark{2}
}

\authorblockA{
\authorrefmark{1}
Faculty of Engineering and Information Technology, University of Technology Sydney\\
Sydney, Australia \\
E-mail: Harunori.Kawano@student.uts.edu.au}

\authorblockA{
\authorrefmark{2}
Dept. of Electrical Engineering and Computer Science, Shibaura Institute of Technology\\
Tokyo, Japan \\
E-mail: sasaki-t@ieee.org}
}

\maketitle
\thispagestyle{empty}
\pagestyle{empty}

\begin{abstract}
While self-supervised learning (SSL) has revolutionized audio representation, excessive parameterization and quadratic computational cost of standard Transformers limit their deployment on resource-constrained devices. To address this bottleneck, we propose HEAR (Human-inspired Efficient Audio Representation), a novel decoupled architecture. Inspired by the human cognitive ability to isolate local acoustic features from global context, HEAR splits the processing pipeline into two dedicated modules: an Acoustic Model for local feature extraction and a Task Model for global semantic integration. Coupled with an Acoustic Tokenizer trained via knowledge distillation, our approach enables robust Masked Audio Modeling (MAM). Extensive experiments demonstrate that HEAR requires only 15M parameters and 9.47 GFLOPs for inference, operating at a fraction of the computational cost of conventional foundation models (which typically require 85M--94M parameters). Despite this high efficiency, HEAR achieves highly competitive performance across diverse audio classification benchmarks. The code and pre-trained models are available at \url{https://github.com/HarunoriKawano/HEAR}.
\end{abstract}

\section{Introduction}
Self-supervised learning (SSL) has revolutionized representation learning in computer vision \cite{DINO, bao2022beit, mae} and natural language processing \cite{devlin2019bert, roberta, clark2020electra}. Building on these foundational breakthroughs, SSL has emerged as the standard paradigm for audio representation, effectively mitigating the reliance on large-scale labeled data. Following pioneering models \cite{hsu2021hubert,wavlm} like wav2vec 2.0 \cite{baevski2020wav2vec2} that established the effectiveness of SSL primarily for speech recognition, spectrogram-based architectures \cite{huang2022audiomae,baade2022maeast,chen2023beats} such as AST \cite{ast} and SSAST \cite{gong2022ssast} have expanded this success to diverse downstream tasks.

While these pre-training paradigms achieve excellent recognition accuracy across diverse tasks, they substantially increase model sizes and computational cost. This creates a critical bottleneck for on-device execution and real-time processing under limited resources. A primary cause of this excessive parameterization and computational overhead is the homogeneous scaling of standard Transformer architectures \cite{vaswani2017attention}. To capture both fine-grained acoustic details and high-level semantics simultaneously, existing foundation models naively stack deep, heavily parameterized layers. This monolithic approach inherently leads to massive model sizes. Furthermore, their self-attention mechanism scales quadratically ($O(N^2)$) with the input sequence length $N$. This structural limitation compounds the computational burden, presenting a major technical barrier for processing long and continuous audio inputs.

To overcome this computational cost, we draw inspiration from the human auditory system. Rather than processing continuous waveforms entirely as global information, human hearing develops through two distinct stages: an innate capability to extract localized acoustic features, and a subsequently acquired ability to interpret high-level, context-dependent semantics. Building on this cognitive structure, we propose HEAR (Human-inspired Efficient Audio Representation), a novel framework that significantly reduces computational costs while maintaining high representational capacity. HEAR partitions the architecture into two modules: an Acoustic Model that focuses exclusively on extracting local acoustic properties, and a Task Model that integrates global context for specific downstream tasks. This decoupling not only circumvents the quadratic bottleneck by restricting the initial processing to local features, but also prevents model enlargement by allowing each specialized module to remain highly compact.

Furthermore, we incorporate the concept of information compression, mirroring how humans convert continuous acoustic signals into discrete semantic units like phonemes. By employing SSL that targets semantically condensed discrete labels rather than a continuous feature space, our Acoustic Model robustly acquires underlying sound structures without overfitting to superficial noise.

\begin{figure*}[t]
    \centering
    \includegraphics[width=0.75\textwidth]{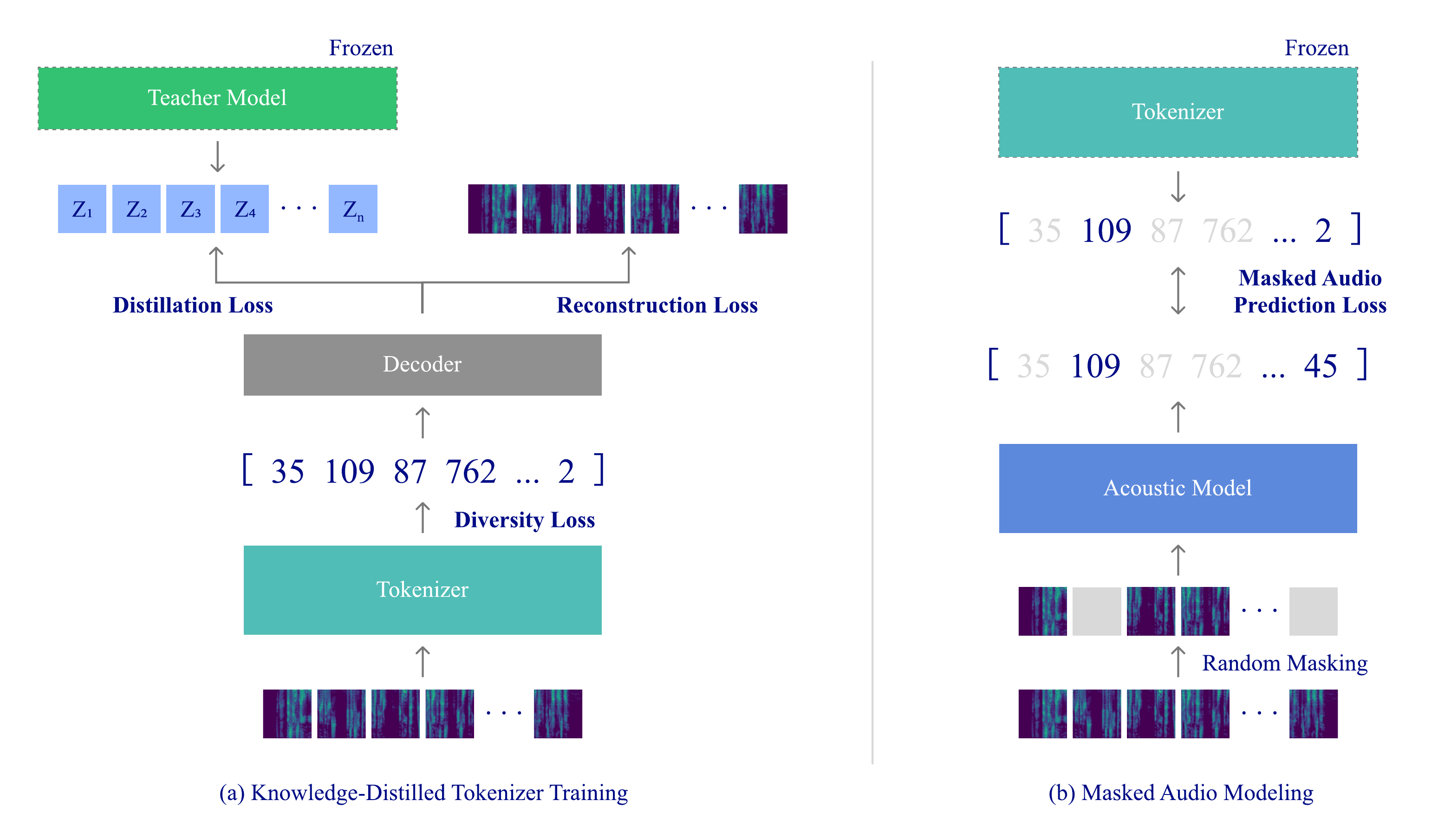} 
    \caption{Overview of the pre-training pipeline for the proposed HEAR framework. (a) Knowledge-Distilled Tokenizer Training: An acoustic tokenizer is trained to encode continuous mel-spectrogram patches into discrete semantic tokens, optimized via signal reconstruction, codebook diversity, and knowledge distillation from a pre-trained teacher model. (b) MAM: Using the frozen tokenizer, the Acoustic Model is pre-trained to predict the discrete tokens of independent random masked patches, enabling robust local feature extraction.}
    \label{fig:pre-training}
\end{figure*}

The main contributions of this work are summarized as follows:
\begin{itemize}
    \item We propose HEAR, a computationally efficient architecture inspired by the human auditory system that structurally decouples local feature extraction (Acoustic Model) and global task adaptation (Task Model).
    \item We introduce an Acoustic Tokenizer based on Gumbel-Softmax \cite{jang2017gumbel} and knowledge distillation \cite{hinton2015distilling} to emulate the discrete interpretation of audio signals, enabling robust Masked Audio Modeling (MAM).
    \item We demonstrate that HEAR requires only 15M parameters and 9.47 GFLOPs, operating at a fraction of the computational cost of existing foundation models (typically 85M--94M parameters). Despite this highly compact size, HEAR maintains comparable accuracy across multiple audio classification tasks.
\end{itemize}

\section{Methods}
\subsection{Architecture Overview}
The architecture of HEAR is grounded in the decoupled cognitive processing of the human auditory system. Specifically, it comprises two primary modules: an Acoustic Model responsible for extracting local acoustic features, and a Task Model that integrates these feature sequences to adapt to specific downstream tasks.

In the initial inference stage, the input audio waveform is converted into a log-mel spectrogram and normalized using instance-specific mean and standard deviation. Subsequently, we apply patch splitting along the time axis with a kernel size of 2 and a stride of 2, yielding a 1D input token sequence. To avoid the quadratic computational complexity ($O(N^2)$) bottleneck with respect to sequence length, the Acoustic Model divides the input audio into fixed-length local chunks of 6 seconds and processes each independently. Because the relative distance and sequential relationships between adjacent acoustic events play an essential role in forming local patterns, we employ a Transformer encoder incorporating relative position representations \cite{relative_position} as the backbone of the Acoustic Model. 

The sequence of local features extracted by the Acoustic Model is ultimately fed into the Task Model. To strictly validate the core concept of decoupling feature extraction from task adaptation, we employ a standard Transformer architecture for the Task Model. The Task Model applies a global self-attention mechanism to the input local feature sequence, integrating the global context across the entire sequence. This maps the universal acoustic representations acquired by the Acoustic Model into the high-level semantics required for specific downstream tasks, yielding the final prediction.

To fully exploit the potential of the proposed decoupled architecture, we design a three-stage training pipeline:
\begin{enumerate}
    \item \textbf{Acoustic Tokenizer Training:} Acquiring high-quality discrete representations from continuous audio via signal reconstruction and knowledge distillation.
    \item \textbf{Acoustic Model Pre-training:} Learning local feature extraction capabilities through MAM, which predicts the discrete labels generated by the tokenizer.
    \item \textbf{Downstream Adaptation:} Combining the pre-trained Acoustic Model with the Task Model to perform fine-tuning via a feature fusion process.
\end{enumerate}
We detail each phase below.

\subsection{Knowledge-Distilled Tokenizer Training}
The first stage of our proposed framework, Acoustic Tokenizer pre-training, aims to extract semantically rich discrete representations from continuous audio signals. As illustrated in Fig.~\ref{fig:pre-training}(a), this framework builds upon the VQ-VAE architecture \cite{oord2017vqvae}. It consists of a Tokenizer that encodes input signals into latent representations and a Decoder that reconstructs the original acoustic features. Similar to the Acoustic Model, both the Tokenizer and Decoder employ Transformers with relative position representations, designed with smaller parameter sizes to ensure training efficiency.

The Tokenizer encodes the input audio signal and quantizes it into a 32-dimensional codebook space with a vocabulary size of $V=1024$. To generate discrete labels in a differentiable manner, we employ the Gumbel-Softmax operation. The temperature parameter $\tau$ starts at $2.0$ and is exponentially annealed by a factor of $0.999995$ at each training step until it reaches $0.5$. This schedule facilitates a smooth transition from diverse codebook exploration to strict discrete labels.

The fundamental objective is to accurately reconstruct the original audio signal. We define the reconstruction loss ($L_{rec}$) as the mean squared error between the input sequence of normalized log-mel spectrogram patches and the reconstructed sequence generated by the Decoder. Minimizing this loss preserves the foundational acoustic properties.

To mitigate codebook collapse, a common issue where the model utilizes only a small subset of codebook vectors, we introduce a diversity loss ($L_{div}$). This loss penalizes the mean squared deviation of the batch-wise codebook assignment probabilities from a uniform distribution.

Furthermore, we incorporate knowledge distillation to efficiently guide the Tokenizer training. We compute a distillation loss ($L_{distill}$) based on the cosine similarity between the Decoder's output representations and the high-quality latent features extracted by a pre-trained teacher model. This enables the Tokenizer to inherit structural acoustic semantics beyond superficial signal reconstruction.

The overall Tokenizer training optimizes a combined objective function, parameterized by weighting coefficients $\alpha$, $\beta$, and $\gamma$:
\begin{equation}
L_{total} = \alpha L_{rec} + \beta L_{distill} + \gamma L_{div}
\end{equation}

\subsection{Masked Audio Modeling}
For the second stage of our proposed framework, the SSL of the Acoustic Model, we adopt MAM following the BEiT paradigm \cite{bao2022beit}. In this training process (Fig.~\ref{fig:pre-training} (b)), the weights of the Tokenizer pre-trained in the first stage are strictly frozen, serving solely to generate discrete acoustic tokens from the input audio as training targets.

Specifically, the frozen Tokenizer quantizes the input log-mel spectrogram patches into a discrete acoustic token sequence $Z = (z_1, \dots, z_N)$. We then define a subset of randomly selected indices $M$ to construct a partially corrupted input sequence by replacing the values of the patches at these positions with zeros. We apply independent random masking to each patch, which naturally encourages the Acoustic Model to focus on robust local feature extraction. This differs from the continuous span masking standardly used in models like wav2vec 2.0.

The Acoustic Model predicts the original discrete tokens at the corrupted positions, optimized via cross-entropy loss (Masked Audio Prediction loss) against the ground-truth tokens. By optimizing this loss, the model efficiently acquires the semantic and universal acoustic structures underlying the audio signal, without overfitting to superficial continuous-value fluctuations or noise.

\begin{figure}[tb]
\centering
\includegraphics[width=0.75\columnwidth]{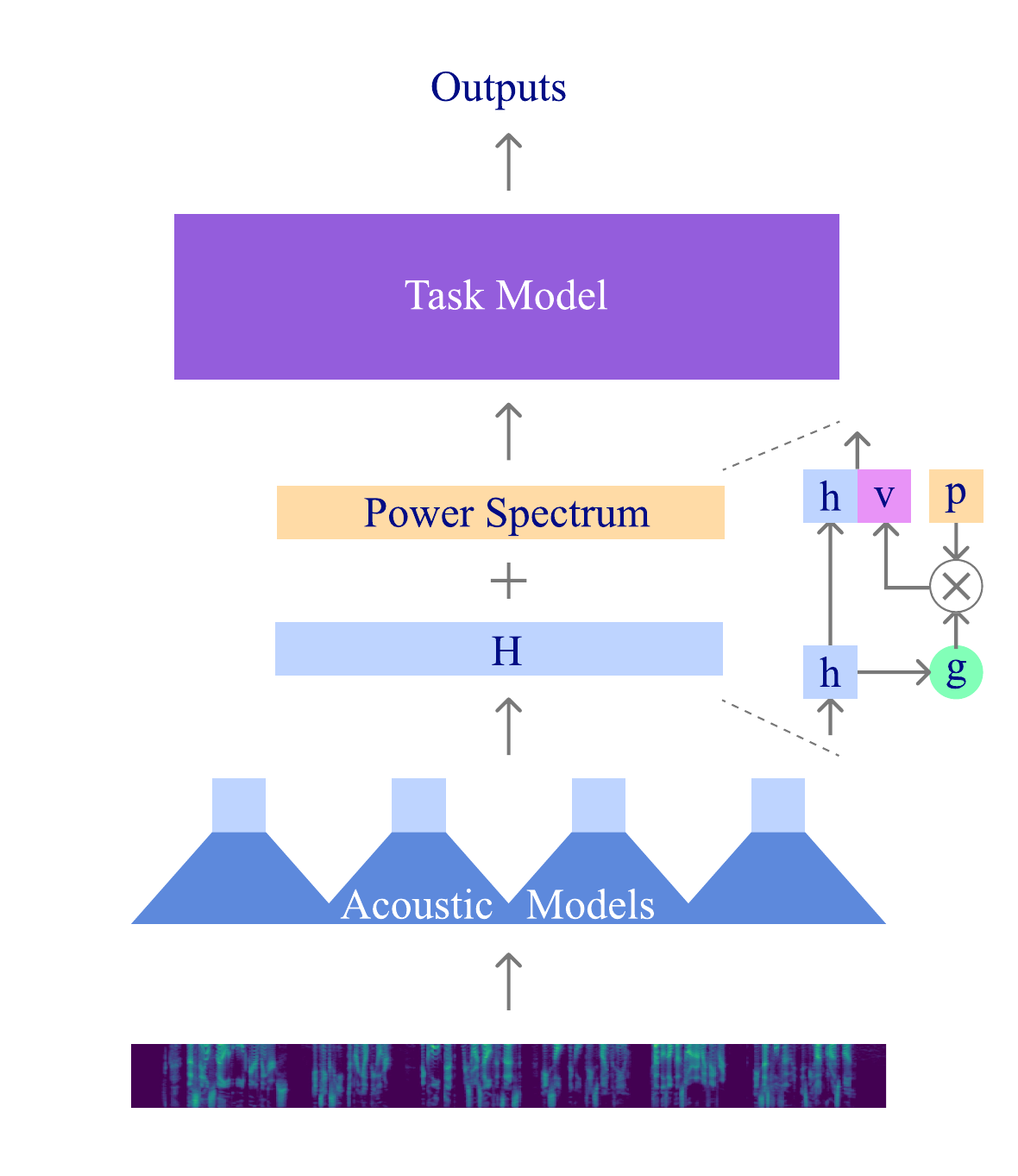} 
\caption{Comprehensive flow of the downstream adaptation, illustrating the feature gating integration of the Acoustic Model outputs and raw power spectrum into the Task Model.}
\label{fig:downstream_tasks}
\end{figure}

\subsection{Downstream Adaptation with Feature Gating}In the final phase, we integrate the pre-trained Acoustic Model with a Task Model for fine-tuning on specific downstream tasks (Fig.~\ref{fig:downstream_tasks}).

Real-world audio inputs often exceed the fixed input length of the model. To process variable-length audio, we apply a sliding window approach, dividing the continuous audio into 6-second segments with a 1-second overlap. The Acoustic Model processes each segment independently. To prevent feature discontinuities at segment boundaries, we apply a cosine-based cross-fade merge within the overlap region. This operation yields a continuous, unified sequence of local acoustic features, where the representation at time step $t$ is denoted as $\bar{h}_t$.

While the Acoustic Model captures robust latent representations, explicit physical acoustic properties are often critical for specific tasks. To incorporate these properties, we introduce a feature gating mechanism that fuses the z-normed log-power spectrum with the Acoustic Model output. Let $S_t$ be the raw power spectrum input at time $t$. We first project $S_t$ into a $d_p$-dimensional vector $v_t \in \mathbb{R}^{d_p}$ using a linear layer with learnable weights $W_s$ and bias $b_s$:
\begin{equation}
v_t = W_s S_t + b_s
\end{equation}
Concurrently, we compute a gate vector $g_t \in \mathbb{R}^{d_p}$ based on the acoustic feature $\bar{h}_t$ using a linear transformation followed by a sigmoid function $\sigma$:
\begin{equation}
g_t = \sigma(W_g \bar{h}_t + b_g)
\end{equation}
We then apply element-wise multiplication between the gate vector and the projected spectrum to extract a task-specific representation $\tilde{v}_t$. This operation selectively emphasizes critical frequency bands while suppressing irrelevant information:
\begin{equation}
  \tilde{v}_t = g_t \odot v_t  
\end{equation}
Finally, the gated physical feature $\tilde{v}_t$ is concatenated with the acoustic representation $\bar{h}_t$ to form the integrated input $c_t = [\bar{h}_t ; \tilde{v}_t]$. 

The Task Model receives the full sequence $C = (c_1, \dots, c_T)$ and processes it through a single-layer Transformer to capture global dependencies. To derive a fixed-size global representation, we compress the resulting sequence along the time dimension by calculating its mean, standard deviation, and maximum values. These three pooled features are then concatenated and fed into a fully connected layer to perform the final classification tailored to the downstream task.

\section{Experiments}
% TODO: Write experiments (~2.0 pages)

\subsection{Experimental Setup}

\subsubsection{Pre-training}
To acquire versatile acoustic representations, we constructed a large-scale and diverse pre-training dataset totaling over 10,000 hours by integrating three datasets: AudioSet \cite{audioset}, VGG-Sound \cite{chen2020vggsound}, and LAION-Audio-630k \cite{LAION-dataset}. All audio data was resampled to 16 kHz and segmented into blocks with a maximum duration of 6 seconds. To maintain the consistency and information density of temporal features, segments shorter than 1 second were excluded from the training set. Both the Tokenizer and the Acoustic Model were pre-trained on this identical integrated dataset. The detailed architectural parameters of each model are summarized in Table~\ref{tab:architecture}.

During the Acoustic Tokenizer training phase, we adopted a pre-trained SSAST as the teacher model for knowledge distillation. We optimized the model using AdamW with a peak learning rate of $3 \times 10^{-4}$ and a batch size of 64 for 540K steps. The learning rate schedule consisted of a linear warm-up to the peak value over the first 10,000 steps, followed by cosine decay. The loss weighting coefficients were empirically set to $\alpha = 0.3$ (reconstruction loss), $\beta = 0.7$ (distillation loss), and $\gamma = 0.1$ (diversity loss).

For the Acoustic Model pre-training, we strictly froze the weights of the pre-trained Tokenizer and performed MAM with a random masking ratio of 40\%. Optimization was similarly conducted using AdamW with a peak learning rate of $5 \times 10^{-4}$ and a batch size of 128 for 450K steps. Consistent with the Tokenizer training, we applied a learning rate scheduler featuring a 10,000-step linear warm-up followed by cosine decay.

\subsubsection{Downstream Tasks}
To evaluate the universality and transferability of the learned acoustic representations, we fine-tuned our model on multiple downstream tasks with diverse acoustic characteristics and requirements. We utilized the following four standard benchmark datasets:

\begin{itemize}
    \item \textbf{ESC-50} \cite{piczak2015esc}: A dataset for environmental sound classification consisting of 50 classes. It contains 2,000 audio clips, each 5 seconds long. We evaluated the performance following the official 5-fold cross-validation protocol.
    \item \textbf{Speech Commands v1 (GSCv1)} \cite{gsc}: A dataset for keyword spotting. We formulated this as a 12-class classification task, comprising 10 specific keywords, 1 silence class, and 1 unknown class containing 20 other common speech commands. For testing accuracy, we used the data and splits provided by the SUPERB benchmark \cite{superb}.
    \item \textbf{Speech Commands v2 (GSCv2)} \cite{gsc}: An extension of GSCv1 containing 35 classes of speech commands. We evaluated this as a 35-class classification task covering all target words.
    \item \textbf{VoxCeleb} \cite{nagrani2017voxceleb}: A large-scale speaker identification dataset recorded in real-world environments. We used this as a multi-class classification task to evaluate how robustly the model extracts speaker-specific acoustic characteristics independent of spoken content and background noise.
\end{itemize}

For all downstream tasks, we adopted accuracy as the evaluation metric. The detailed hyperparameter settings for fine-tuning on each task are summarized in Table~\ref{tab:finetune_params}.

\begin{table}[t]
\begin{center}
\begin{threeparttable}
\caption{Detailed architectural hyperparameters for each module in the proposed framework.}
\label{tab:architecture}
\footnotesize
\setlength{\tabcolsep}{3pt}
\begin{tabular}{l c c c c c}
\toprule
\textbf{Module} & \textbf{Hidden Size} & \textbf{Intermediate Size} & \textbf{Heads} & \textbf{Layers} \\
\midrule
Tokenizer      & 128 & 512  & 4 & 6 \\
Decoder        & 128 & 512  & 4 & 2 \\
Acoustic Model & 384 & 1536 & 4 & 6 \\
Task Model     & 384 & 1536 & 4 & 1 \\
\bottomrule
\end{tabular}
\end{threeparttable}
\end{center}
\end{table}

\begin{table}[tb]
\begin{center}
\begin{threeparttable}
\caption{Hyperparameter settings for fine-tuning on downstream tasks.}
\label{tab:finetune_params}
\footnotesize
\setlength{\tabcolsep}{8pt}
\begin{tabular}{l c c c}
\toprule
\textbf{Dataset} & \textbf{Batch Size} & \textbf{Learning Rate} & \textbf{Max Epochs} \\
\midrule
ESC-50   & 16 & $2 \times 10^{-4}$ & 80 \\
GSCv1    & 64 & $3 \times 10^{-4}$ & 30 \\
GSCv2    & 64 & $3 \times 10^{-4}$ & 30 \\
VoxCeleb & 16 & $1 \times 10^{-4}$ & 15 \\
\bottomrule
\end{tabular}
\end{threeparttable}
\end{center}
\end{table}

\subsection{Performance Comparison}
Table~\ref{tab:performance_comparison} compares the performance and inference efficiency of HEAR against representative baseline models. The primary advantage of HEAR is its exceptional computational efficiency. It requires only 15M parameters in total, which explicitly consists of a 12M Acoustic Model, a 2M Task Model, and a 1M fully connected classification head. With this highly compact design, it is approximately 16\% the size of models like wav2vec 2.0 (94M). Furthermore, HEAR achieves significant speedups and reductions in computational cost, recording 9.47 GFLOPs for a 10-second audio input and a Real-Time Factor (RTF) of 0.095, substantially outperforming the baselines.

\begin{table*}[t]
\begin{center}
\begin{threeparttable}
\caption{Performance and inference efficiency comparison between HEAR and representative baseline models.}
\label{tab:performance_comparison}
\footnotesize
\begin{tabular}{l c c c c c c c}
\toprule
\multirow{2}{*}{\textbf{Model}} & \multicolumn{4}{c}{\textbf{Accuracy (\%)}} & \multicolumn{3}{c}{\textbf{Efficiency}} \\
\cmidrule(lr){2-5} \cmidrule(lr){6-8}
& \textbf{ESC-50} & \textbf{GSCv1} & \textbf{GSCv2} & \textbf{VoxCeleb} & \textbf{Params} & \textbf{GFLOPs}\tnote{a} & \textbf{RTF}\tnote{a} \\
\midrule
wav2vec 2.0 \cite{baevski2020wav2vec2} & -    & 96.2\tnote{b} & -    & 75.1\tnote{b} & 94M & 69.6 & 0.510 \\
HuBERT \cite{hsu2021hubert}      & -    & 96.3\tnote{b} & -    & 81.4\tnote{b} & 94M & 69.6 & 0.510 \\
AudioMAE \cite{huang2022audiomae}  & 94.1 & 96.9 & 98.3 & 94.8 & 85M & 42.4 & 0.244 \\
SSAST \cite{gong2022ssast}       & 88.8 & 96.0 & 98.0 & 64.2 & 85M & 42.4 & 0.244 \\
BEATs \cite{chen2023beats}       & 95.6 & 97.7 & 98.3 & -    & 85M & 42.4 & 0.244 \\
\textbf{HEAR (Ours)} & 84.9 & 94.3 & 95.1 & 87.9 & \textbf{15M} & \textbf{9.47} & \textbf{0.095} \\
\bottomrule
\end{tabular}
\begin{tablenotes}
\item[a] To simulate an edge device environment, inference efficiency (GFLOPs and RTF for a 10-second audio input at a batch size of 1) was measured using a single thread of an ARM-based processor with 4GB of RAM (Google Cloud).
\item[b] Results are from the SUPERB benchmark \cite{superb} (frozen pre-trained model).
\end{tablenotes}
\end{threeparttable}
\end{center}
\end{table*}

Despite this drastic model compression, HEAR maintains highly competitive performance across various tasks. It achieves accuracies of 94.3\% on GSCv1, 95.1\% on GSCv2, and 84.9\% on ESC-50. Compared to state-of-the-art (SOTA) models that are over six times larger, HEAR exhibits only marginal performance drops, demonstrating an excellent accuracy-efficiency trade-off.

In speaker identification (VoxCeleb), HEAR achieves an accuracy of 87.9\% under the full fine-tuning setting. For a more aligned comparison with the SUPERB baselines, which evaluate frozen features, the transfer setting in our ablation study (Table~\ref{tab:ablation}) demonstrates that our frozen Acoustic Model achieves 76.0\%, performing comparably to the 94M-parameter wav2vec 2.0 (75.1\%).

\subsection{Effectiveness of Components}
To validate the effectiveness of each component in the proposed method, we conducted an ablation study focusing on three aspects: the contribution of SSL, the efficacy of the gated power spectrum input, and a comparison between freezing the Acoustic Model weights and full fine-tuning. The results of these comparative experiments are summarized in Table~\ref{tab:ablation}.

\subsubsection{Impact of Pre-training}
When training from scratch with random weight initialization without pre-training (Scratch), accuracy degraded significantly across all tasks. This performance drop is particularly pronounced on ESC-50, which has a relatively small dataset size (dropping from 84.9\% to 62.1\%), and VoxCeleb, which requires complex feature extraction (dropping from 87.9\% to 72.2\%). This demonstrates that large-scale SSL via MAM is essential for acquiring versatile and robust acoustic representations.

\subsubsection{Gated Power Spectrum Analysis}
Removing the explicit power spectrum input resulted in a slight accuracy drop on ESC-50 (84.7\%) and VoxCeleb (87.6\%). Conversely, performance marginally improved on speech command tasks like GSCv1 (from 94.3\% to 94.7\%). These results suggest that while explicitly providing spectral information is effective for tasks heavily reliant on physical acoustic characteristics (e.g., environmental sound classification and speaker identification), it is not necessarily required for tasks primarily focused on recognizing spoken content.

\subsubsection{Frozen vs. Fine-tuned}
Completely freezing the Acoustic Model weights and training only the Task Model (Transfer) yielded contrasting results depending on the nature of the task. For GSCv1 (95.1\%) and GSCv2 (95.8\%), this approach outperformed the full model fine-tuning (Base: 94.3\% and 95.1\%, respectively). This is likely because the feature space acquired during pre-training is already highly adapted to speech recognition tasks, and freezing the weights helped prevent overfitting. In contrast, accuracy dropped significantly on ESC-50 (78.5\%) and VoxCeleb (76.0\%), indicating that deep adaptation (fine-tuning) of the entire Acoustic Model to the target task is crucial for identifying complex environmental sounds and speaker characteristics.

\begin{table}[tb]
\begin{center}
\begin{threeparttable}
\caption{Ablation study results (Accuracy \%) on downstream tasks.}
\label{tab:ablation}
\footnotesize
\setlength{\tabcolsep}{4pt}
\begin{tabular}{l c c c c}
\toprule
\textbf{Setting} & \textbf{ESC-50} & \textbf{GSCv1} & \textbf{GSCv2} & \textbf{VoxCeleb} \\
\midrule
Base             & \textbf{84.9} & 94.3 & 95.1 & \textbf{87.9} \\
Scratch          & 62.1 & 88.1 & 89.5 & 72.2 \\
w/o Spectrum     & 84.7 & 94.7 & 95.0 & 87.6 \\
Transfer         & 78.5 & \textbf{95.1} & \textbf{95.8} & 76.0 \\
\bottomrule
\end{tabular}
\end{threeparttable}
\end{center}
\end{table}

\section{Conclusion}
In this paper, we proposed HEAR, a human-inspired audio representation framework that structurally decouples local feature extraction from global task adaptation. This separation effectively resolves the quadratic computational bottleneck of standard Transformers. Requiring only 15M parameters and 9.47 GFLOPs, HEAR operates at a fraction of the cost of existing foundation models while maintaining competitive accuracy across diverse downstream tasks. This efficiency holds significant promise for real-time edge computing, with future work targeting on-device deployment and complex sequence transduction tasks like automatic speech recognition.

\section*{Acknowledgment}
The authors would like to thank the members of the laboratory at Shibaura Institute of Technology for their technical support and constructive discussions.
\newpage

\printbibliography

\end{document}